# Previous R&D of vibrating wire alignment technique for HEPS

WU Lei(吴蕾)[1,2]    WANG Xiao-long(王小龙)[1,3]    LI Chun-hua(李春华)[1]    QU Hua-min(屈化民)[1,3]

[1] Institute of High Energy Physics, Chinese Academy of Sciences, Beijing 100049, China

[2] University of Chinese Academy of Sciences, Beijing 100049, China

[3] Dongguan Institute of Neutron Science (DINS), Dongguan 523808, China

**Abstract:**   The alignment tolerance of multipoles on a girder is better than ±30μm in the storage ring of High Energy Photon Source (HEPS) which will be the next project at IHEP (Institute of High Energy Physics). This is difficult to meet the precision only using the traditional optical survey method. In order to achieve this goal, vibrating wire alignment technique with high precision and sensitivity is considered to be used in this project. This paper presents some previous research works about theory, scheme design and achievements.

## 1   Introduction

HEPS will be a 5 GeV, 1296 meters circumference third generation synchrotron radiation facility with ultra emittance and extremely high brightness. The emittance will be better than 0.1nm·rad. The storage ring is a 48 cell 7BA lattice. Figure 1 is one of 48 typical cells. The multipole girder that supports several quadrupoles and sextuples is designed 3.8 meters. But this is not the final value, the lattice is still in the design [1].

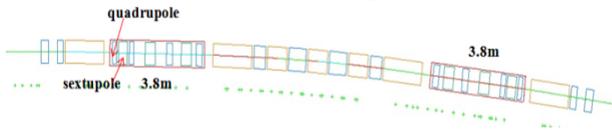

Fig. 1.   One of 48 typical cells.

Table 1.   HEPS alignment tolerance.

| Tolerances | Magnet to Magnet | Girder to Girder |
|---|---|---|
| Horizontal | ±0.03mm | ±0.05mm |
| vertical | ±0.03mm | ±0.05mm |
| Beam direction | ±0.5mm | ±0.5mm |
| Roll angle | ±0.2mrad | ±0.2mrad |

Table 1 shows the alignment tolerance in HEPS. It is difficult to achieve the required accuracy using the traditional optical survey. So vibrating wire alignment technique is considered to meet the tolerance ±0.03mm between magnet to magnet on one multipole girder. Besides, automatic adjustment girder can help to achieve the tolerance ±0.05mm between girder to girder.

Vibrating wire technique has been used in some different projects in many labs. It has demonstrated the potential to measure the magnetic center to the required accuracy. These applications of vibrating wire are based on the same fundamental principle, but the specific purposes are distinct. Vibrating wire is firstly be used at Cornell University to find the magnetic center of superconducting quadrupoles placed inside the cryostat for CESR [2]. Later, it has been used to find the solenoid magnetic center [3] and study the characterization of undulator magnets [4]. In SLAC, this technique is used to fiducialize the quadrupoles between undulator segments for LCLS [5] and align quadrupole for FFTB. In CERN, its application is solenoid magnetic center finding [6]. In PSI, vibrating wire is used to find the magnetic axis of quadrupoles for Swiss Free Electron Laser [7]. In BNL, this technique is used to align the quadrupoles and sextuples on one girder for NSLS-II [8].

Vibrating wire is based on measurement of the magnetic axis to align the magnets. It is a further evolution of pulsed wire method. The simple diagrammatic sketch is like Figure 2 [9]. The specific principle of this method is like that: a single conducting wire is stretched through the magnet aperture and driven by an alternating current. Transversal vibrations are continuously excited by period Lorentz Force. By matching current frequency to one of the resonant modes of wire, the vibration amplitude and sensitivity are enhanced and the magnetic induction intensity at the wire position can be calculated. Move the wire across the aperture in horizontal direction(x) or vertical(y) direction, the magnetic induction intensity distribution can be gotten and the magnetic axis can be found.

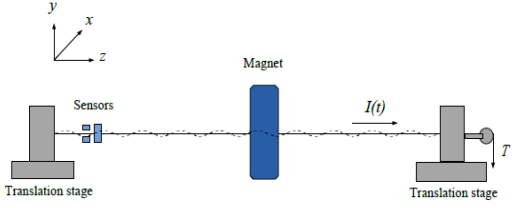

Fig. 2. Diagrammatic sketch of vibrating wire.

# 2 Research development

So far, the main research development is related to the research work about theory and scheme design, including derivation of basic theory, design of test bench model, design of sag measurement scheme, calculation of vibrating amplitudes, design of sensor circuit and sensor calibration and design of data acquisition and control system.

## 2.1 Basic theory

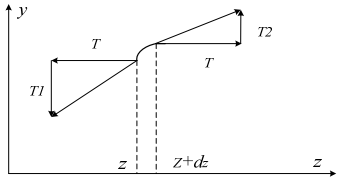

Fig. 3. Force analysis of a section of wire $dz$.

According to Newton's law, the differential equation of the motion of a section of wire $dz$ in y-z plane (Figure 3) is:

$$\mu \frac{\partial^2 y}{\partial t^2} = T \frac{\partial^2 y}{\partial z^2} - \mu g - \gamma \frac{\partial y}{\partial t} + B_x(z) I_0 \exp(i\omega t). \quad (1)$$

$\mu$ is the mass per unit length. T is the wire tension. $\gamma$ is the damping coefficient. Solving the differential equation, the homogenous solution is:

$$\begin{aligned} y_h(z,t) = \mathrm{Re} \sum_{n=1}^{\infty} \sin(\frac{n\pi}{l} z) &[A e^{-\frac{\alpha}{2} t} \cos \sqrt{\omega_n^2 - (\frac{\alpha}{2})^2} t \\ &+ B e^{-\frac{\alpha}{2} t} \sin \sqrt{\omega_n^2 - (\frac{\alpha}{2})^2} t )]. \end{aligned} \quad (2)$$

The particular solution due to gravity is:

$$y_g(z) = \frac{\mu g}{2T} z(z-l) = \frac{g}{8 f_1^2 l^2} z(z-l). \quad (3)$$

The particular solution due to Lorentz Force is:

$$\begin{aligned} y_d(z,t) &= Y_B(z) \exp(i\omega t) \\ &= \mathrm{Re} \sum_{n=1}^{\infty} \frac{-I_0 B_{sn}}{\mu(\omega^2 - \omega_n^2 - i\omega\alpha)} \sin(\frac{n\pi z}{l}) \exp(i\omega t). \end{aligned} \quad (4)$$

## 2.2 Vibrating wire test bench model

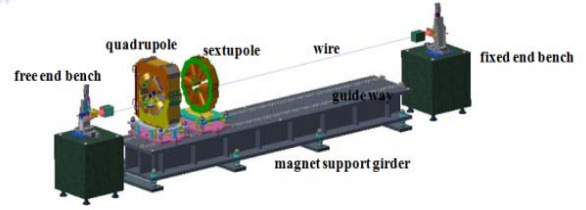

Fig. 4. Vibrating wire test bench model.

Figure 4 is the design of vibrating wire test bench model. The whole test bench is composed of two parts which are end bench and magnet support girder. The difference between fixed end bench and free end bench is the fixed way of the wire. At the fixed end, the wire is fixed and located by V notch. At the free end, the wire is pulled by weights through a pulley and also located by V notch. On each end bench, there are two translation stages and two sensors. The horizontal stage moves wire in x direction and the vertical one for wire movement in y direction. The horizontal sensor detects the wire vibration amplitudes in x direction and the vertical one for y direction. Vibrating wire is 0.125mm diameter wire made with alloy of copper-beryllium. The magnet support girder is designed 5.4 meters. Two parallel guideways are installed above the girder in order to move magnets conveniently. By move magnets along the guideways, the magnetic center could be detected in different place along the girder. The quadrupole and sextupole are borrowed from BEPCII for lack of magnets made for HEPS. They are 105Q and 130S.

## 2.3 Sag measurement scheme

For a 7 meters long wire, sag is unavoidable because of self-weight. It is about several hundred microns. To find the vertical magnetic center, accurate sag correction is essential. Figure 5 is the sag measurement scheme.

The method is to establish a coordinate system relative to the wire and then measure the sag at different place along the wire (z direction). Use laser tracker and NIVEL200 to adjust the position and orientation of the sensor. Make the sensor the same attitudes in different place. Use sensor to measure the vertical position of the wire.

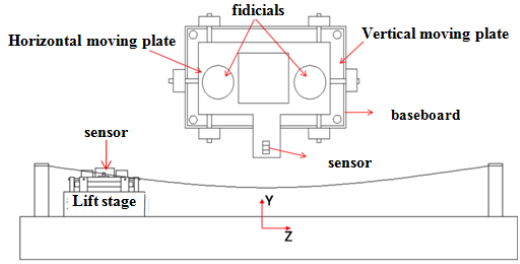

Fig. 5.    Sag measurement scheme.

## 2.4    Sensor circuit and calibration

The product model of the sensor in HEPS is GP1S094HCZ0F. It is a photointerrupter with opposing emitter and detector in a molding that provides non-contact sensing. The function of the sensor is to detect the vibrating amplitudes $y_0(z,t)$ change over time due to Lorentz Force. Because the sensor is just a simple electron component, the sensor circuit is designed to make the output voltage appropriate.

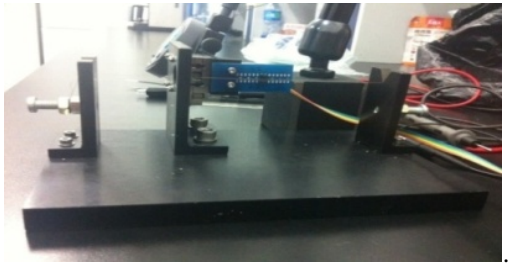

Fig.6.    Sensor output test bench.

We do the sensor calibration on sensor output test bench (Figure 6). By changing the position of the sensor relative to the wire on the sensor output test bench, a series of points which are the output voltages were gotten. Because the sensor received different optical signals when the wire at different locations, so the output voltages are different. Figure 7 and 8 is the outputs of one of four sensors that were tested. The results were summarized in Table 2 and 3. There are two linear output parts. In each part, the wire displacement is about 0.1~0.12mm. One part will be chosen as the workspace. There was little difference in the output voltages between covering the sensor with light shield and exposing the sensor in the lab light. But the voltages were more stable when the sensor was covered. The sensitivity of the sensor is approximately 30mV/μm. In one test position, the measures of variation reduced from 20mV to 6mV after filtering. The stability has improved significantly. So adding filters to the circuit and DAQ program is under consideration.

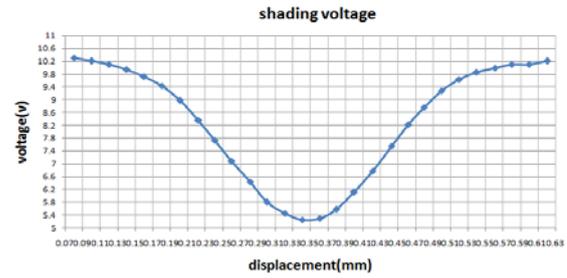

Fig. 7.:    Shading Voltage of Sensor.

Table 2.    Shading voltages of sensor.

|  | Left linear region | | Right linear region | |
|---|---|---|---|---|
| Voltages(V) | 9.45 | 5.8 | 5.58 | 9.27 |
| Voltage difference(V) | 3.65 | | 3.69 | |
| Displacement (mm) | 0.18 | 0.3 | 0.38 | 0.5 |
| Displacement difference(mm) | 0.12 | | 0.12 | |
| Voltage/displace -ment(mV/μm) | 30.42 | | 30.75 | |

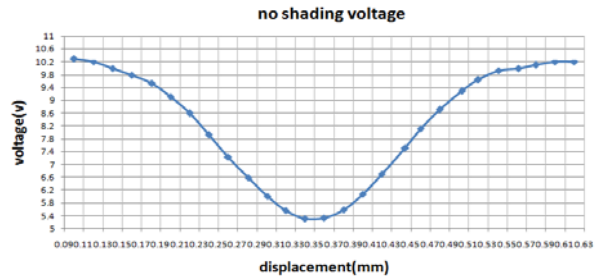

Fig. 8.    No shading voltages of sensor.

Table 3.    No shading voltages of sensor.

|  | Left linear region | | Right linear region | |
|---|---|---|---|---|
| Voltages(V) | 9.1 | 6 | 5.56 | 9.27 |
| Voltage difference(V) | 3.1 | | 3.71 | |
| Displacement (mm) | 0.2 | 0.3 | 0.38 | 0.5 |
| Displacement difference(mm) | 0.1 | | 0.12 | |
| Voltage/displace -ment(mV/μm) | 31 | | 30.92 | |

## 2.5 Calculation of vibrating amplitudes

Vibration amplitudes should be controlled within the voltage linear output part. And it also should be make full use of this region. According to the measurement, the linear region width is about 0.1~0.12mm. We designed the wire vibration amplitudes 0.045mm.

According to equation 4, the equation of measurement of quadrupole magnetic center is deduced:

$$y_{\max}(z) = [\frac{I_0 2Gl_Q}{\mu\omega_n\alpha l}\sin(\frac{n\pi}{j})\sin(\frac{n\pi z}{l})]|\Delta y|. \qquad (5)$$

And we do measurement of sextupole magnetic center using the equation 6.

$$y_{\max}(z) = [\frac{I_0 B''l_Q}{\mu\omega_n\alpha l}\sin(\frac{n\pi}{j})\sin(\frac{n\pi z}{l})]\Delta x^2. \qquad (6)$$

$I_0$ is the wire current amplitude. $Gl_Q$ is the quadrupole integrated gradient. $l_Q$ is the quadrupole length. n is the resonance order. $\omega_n$ is the harmonic frequency. $\alpha$ is the damping constant. J equals $l / z_{magnet}$. $l$ is the length of the wire and $z_{magnet}$ is the location of the magnet along the wire. $B''l_Q$ is the sextupole integrated gradient. $\Delta y$ is the distance between magnetic center relative to the wire in y direction and $\Delta x$ is the distance between magnetic center relative to the wire in x direction.

In short, according to the distance between the magnetic center and wire, the distance between wire end and magnet, the distance between wire end and sensor, and magnetic field gradient, the AC current can be calculated. We use this method to estimate the AC current in the wire according to the previous magnetic survey datum of the magnets borrowed from BEPCII.

## 2.6 Data acquisition and control system

The figure 9 is the scheme of the data acquisition and control system. There are four input signal ranges in DAQ card PXI-6122. The sampling rate is 500Ks per channel. Its function is to sample 2 sensors and 1 driving current signals simultaneously. Datum will be transported to the IPC by the communication cards PXIe- 8360 and PCIe-8362. IPC also controls the movement of 2 x-y stages. We use LabVIEW software to do data acquisition and data process.

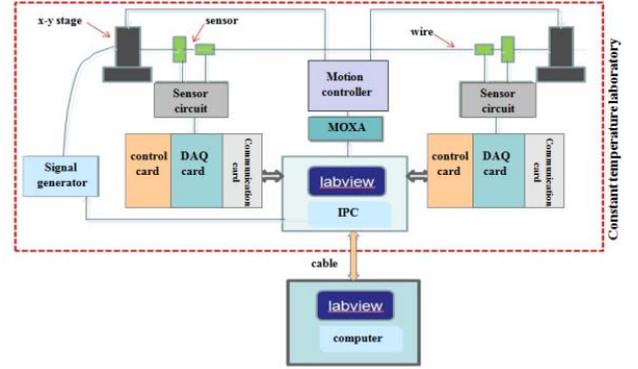

Fig 9. Scheme of data acquisition and control system.

# 3 conclusions

The alignment tolerance of multipoles on a long girder in HEPS should be better than 30 microns. Vibrating wire alignment technique is appropriate to achieve the required accuracy. This technique has never been used in China. Many research works have been done to study the technique. In this article, we introduced the background of the magnets alignment in HEPS and previous research work and achievements.